\documentclass[12pt]{article}

\usepackage{a4}
\usepackage{amsfonts} 
\usepackage{amssymb}
\usepackage{oldgerm}
\usepackage{graphics}

\newcommand{\nit}{\noindent}

\newcommand{\np}{\newpage}
\newcommand{\dsp}{\displaystyle}
\newcommand{\vs}[1]{\vspace{#1 ex}}
\newcommand{\hs}[1]{\hspace{#1 em}}
\newcommand{\bflr}{\begin{flushright}}
\newcommand{\eflr}{\end{flushright}}
\newcommand{\bc}{\begin{center}}
\newcommand{\ec}{\end{center}}
\newcommand{\ben}{\begin{enumerate}}
\newcommand{\een}{\end{enumerate}}

\newcommand{\be}{\begin{equation}}
\newcommand{\ee}{\end{equation}}
\newcommand{\ba}{\begin{array}}
\newcommand{\ea}{\end{array}}
\newcommand{\ct}{\cite}
\newcommand{\bit}{\bibitem}
\newcommand{\dd}[2]{\frac{\partial{#1}}{\partial{#2}}}
\newcommand{\ag}{\alpha}
\newcommand{\bg}{\beta}
\newcommand{\gam}{\gamma}
\newcommand{\del}{\delta}

\newcommand{\ve}{\varepsilon}

\newcommand{\kg}{\kappa}
\newcommand{\lb}{\lambda}
\newcommand{\sg}{\sigma}
\newcommand{\rg}{\rho}

\newcommand{\vf}{\varphi}
\newcommand{\og}{\omega}
\newcommand{\Gam}{\Gamma}

\newcommand{\Og}{\Omega}

\newcommand{\cM}{{\cal M}}

\newcommand{\lh}{\left(}
\newcommand{\rh}{\right)}
\newcommand{\ld}{\left.}
\newcommand{\rd}{\right.}

\newcommand{\der}{\partial}
\newcommand{\nb}{\nabla}

\newcommand{\slashed}{\hspace{-1.1ex}/}
\newcommand{\Slashed}{\hspace{-1.4ex}/\hspace{.2ex}}
\newcommand{\Der}{D}
\newcommand{\sDer}{\Der\Slashed}
\newcommand{\sder}{\der\slashed}

\begin{document}

\pagestyle{empty}

\bflr{Nikhef 2020-024}
\eflr
\vs{5}

\bc
{\bf \large Conformal symmetry and supersymmetry}\\
\vs{2}

{\bf \large in Rindler space}\\
\vs{7}

{\large J.W.\ van Holten} \\
\vs{2} 

Nikhef, Amsterdam NL 
\vs{1} 

and
\vs{1}

Leiden University, Leiden NL
\vs{3}

July 9, 2020
\ec
\vs{5}

\nit
{\small
{\bf Abstract}\\
This paper addresses the fate of extended space-time symmetries, in particular
conformal symmetry and supersymmetry, in two-dimensional Rindler space-time
appropriate to a uniformly accelerated non-inertial frame in flat 1+1-dimensional 
space-time. Generically, in addition to a conformal co-ordinate transformation, the 
transformation of fields from Minkowski to Rindler space is accompanied by local 
conformal and Lorentz transformations of the components, which also affect the 
Bogoliubov transformations between the associated Fock spaces. I construct 
these transformations for massless scalars and spinors, as well as for the ghost 
and super-ghost fields necessary in theories with local conformal and supersymmetries,
as arising from coupling to 2-D gravity or supergravity. Cancellation of the anomalies 
in Minkowski and in Rindler space requires theories with the well-known critical 
spectrum of particles arising in string theory in the limit of infinite strings, 
and is relevant for the equivalence of Minkowski and Rindler frame theories.
}

\np
\pagestyle{plain}
\pagenumbering{arabic}

\nit
{\bf \large 1 Introduction} 
\vs{1}

\nit
Conformal field theories have become an important tool in our theories of nature, both 
in the context of particle physics and quantum gravity and in the context of condensed 
matter \ct{belavin:1984, cardy:1989,schellekens:1995}. Conformal field theories in two 
dimensions are especially relevant in string theory \ct{green:1987}, the AdS$_3$-CFT 
correspondence and condensed matter \ct{fradkin:2013,zaanen:2015}, and as a model 
for quantum field theories also in higher dimensions. 

Effectively conformal field theories in two space-time dimensions can be formulated 
as theories of massless fields in a Minkowski background. This is the point of view 
used for practical purposes in the present paper. In fact our focus is on the description
of these theories in a non-standard frame: that of a uniformly accelerated observer
\ct{rindler:1966}. The interest in this frame stems from the non-inertial effects which
manifest themselves in the presence of a horizon and finite-temperature behaviour: 
the Unruh effect \ct{unruh:1976, davies:1975, wald:1975, fulling:1977}, a counterpart 
of the Hawking effect in black-hole space-times \ct{hawking:1974,hawking:1975} and 
in cosmology \ct{gibbons:1977}. More generally it has been found that free field theories 
in Rindler space describe the near-horizon behaviour of quantum field theories in the 
presence of black holes, which is particularly relevant in the context of AdS-CFT 
correspondence \ct{samantray:2013}. Extensive reviews of field theories in Rindler 
space can be found in refs.\ \ct{brout:1995,crispino:2008}.    

In two space-time dimensions local conformal transformations can be used to 
cast the line element in the form 
\be
ds^2 = \rg^2(x)\, \eta_{\mu\nu} dx^{\mu} dx^{\nu},
\label{1.1}
\ee
where $\eta_{\mu\nu}$ is the flat Minkowski metric and $\rg^2(x)$ is the conformal 
factor, which is constrained by the topology. In the present paper we restrict ourselves 
to inertial observers in infinite flat Minkowski space for which $\rg^2(x) = 1$, and to 
accelerated observers in the associated Rindler space, a subset of Minkowski space 
consisting of two branches: the right and left Rindler wedges $R$ and $L$, defined by
\be
\ba{l}
$R$: \hs{2} x > 0 \hs{1} \mbox{and} \hs{1} - x <  t < x; \\
 \\
$L$:\, \hs{2} x < 0 \hs{1} \mbox{and} \hs{1.3} x < t < -x.
\ea
\label{1.2}
\ee
Each wedge is parametrized by a conformally flat metric in terms of co-ordinates $(\tau, \xi)$
associated with an accelerated observer: 
\be
\ba{ll}
\dsp{ R: \hs{1}  x = \frac{1}{a}\, e^{a\xi} \cosh a \tau, }& \dsp{ t = \frac{1}{a}\, e^{a\xi} \sinh a \tau, }\\
 & \\
\dsp{ L: \hs{1} x = - \frac{1}{a}\, e^{-a\xi} \cosh a \tau, }& \dsp{ t = \frac{1}{a}\, e^{-a\xi} \sinh a \tau. }
\ea
\label{1.3}
\ee
In this parametrization the line element is of the form (\ref{1.1}) with conformal factor
$\rg = e^{\pm a\xi}$:
\be
\ba{l}
R: \hs{1} ds^2 = e^{2a\xi} \lh - d\tau^2 + d\xi^2 \rh, \\ 
 \\
L: \hs{1} ds^2 = e^{-2a\xi} \lh - d\tau^2 + d\xi^2 \rh.
\ea
\label{1.4}
\ee
The parameter $a$ in the co-ordinate transformation represents the proper accelaration 
in the Minkowski frame of a body with Rindler co-ordinate $\xi = 0$ at the moment it 
crosses the $x$-axis. It can be absorbed in a further rescaling of the Rindler co-ordinates 
$(\tau, \xi)$ by $1/a$ and a shift of $\xi$ by $\xi_0 = \ln a$; as a result the line element 
reduces to (\ref{1.4}) with $a = 1$. This is our standard choice in the following. In all of 
this paper units are chosen such that $c = \hbar = 1$.
\vs{2}

\nit
{\bf \large 2.\ Massless fields in 2-D Minkowski space}
\vs{1}

\nit
To fix conventions and for later reference, in this section we briefly review free field 
theories of massless scalar fields (bosons) and massless spinor fields (fermions) in 
2-D Minkowski space. The theory of a real, free scalar field $\vf(x)$ is defined by the 
action
\be
S_b = \frac{1}{2}\, \int_M d^2x \left[ \lh \der_t \vf \rh^2 - \lh \der_x \vf \rh^2 \right].
\label{2.1}
\ee
The solutions of the associated Klein-Gordon equation 
\be
\lh \der^2_t - \der^2_x \rh \vf = 0,
\label{2.2}
\ee
are superpositions of left- and right-moving fields
\be
\vf(t,x) = \vf_+(t+x) + \vf_-(t-x).
\label{2.3}
\ee
For definiteness in the following we concentrate on left-moving fields $\vf_+$. 
Its conjugate field momentum is $\pi_+ = \der_t \vf_+$, and the time-evolution 
of all functionals of the field is governed by the hamiltonian 
\be
H_b = \frac{1}{2}\, \int_{-\infty}^{\infty} dx \left[ \pi_+^2 + (\der_x \vf)^2 \right],
\label{2.4}
\ee
using the equal-time commutation relation 
\be
\left[ \vf_+(t,x), \pi_+(t,y) \right] = i \del(x-y).
\label{2.5}
\ee
In a plane-wave basis the field is expanded as
\be
\vf_+(t+x) = \int_0^{\infty} \frac{dk}{2\pi \sqrt{2k}} \lh a_k e^{-ik(t+x)} + a^*_k e^{ik(t+x)} \rh,
\label{2.6}
\ee
the domain of momenta $k$ for left-moving fields being $\left[0, \infty \rh$. The plane-wave 
mode operators are defined inversely by
\be
a_k = \frac{i}{\sqrt{2k}}\, \int_{-\infty}^{\infty} dx\, e^{ik(t+x)} \stackrel{\leftrightarrow}{\der}_t \vf_+, \hs{2}
a^*_k = - \frac{i}{\sqrt{2k}}\, \int_{-\infty}^{\infty} dx\, e^{-ik(t+x)} \stackrel{\leftrightarrow}{\der}_t \vf_+.
\label{2.7}
\ee
The commutation relation (\ref{2.5}) is equivalent to 
\be
\left[ a_k, a^*_q \right] = 2 \pi \del(k-q),
\label{2.8}
\ee
and the hamiltonian after normal-ordering is 
\be
{:H}_b:\, = \int_0^{\infty} \frac{dk}{2\pi}\, k a^*_k a_k.
\label{2.9}
\ee
Finally the ground state of the field is the bosonic Minkowski vacuum $| 0 \rangle_b$:
\be
\forall k: \hs{1} a_k | 0 \rangle_b = 0.
\label{2.10}
\ee
Next we discuss spinor fields, using the conventions collected in appendix A. A 
general Dirac spinor is a complex 2-component object; however, 2-D Minkowski 
space-time allows Majorana-Weyl spinors which are real single-component objects 
representing left- or right-moving solutions of the massless Dirac equation. Indeed, 
in the Majorana representation making use of the charge conjugation operator
$C$ the Dirac action takes the diagonal form 
\be
S_f = \frac{i}{2}\, \int_M d^2x\, \lb^T C \sder\, \lb, \hs{2}
C \sder =  \lh \ba{cc} \der_t - \der_x & 0 \\ 0 & \der_t + \der_x \ea \rh. 
\label{2.11a}
\ee
and for a 2-component Majorana spinor 
\[
\lb = C \bar{\lb}^T = \left[ \ba{c} \lb_+ \\ \lb_- \ea \right],
\]
the associated Dirac equation splits into independent equations 
\be
\lh \der_t - \der_x \rh \lb_+ = 0, \hs{2} \lh \der_t + \der_x \rh \lb_- = 0,
\label{2.11}
\ee
representing separate left- and right moving fields. Again, for definiteness we 
concentrate on left-moving fields $\lb_+(t+x)$. The time-evolution of functionals of 
this field is governed by the hamiltonian 
\be
H_f = \frac{i}{2}\, \int_{-\infty}^{\infty} dx\, \lb_+ \der_x \lb_+, 
\label{2.13}
\ee
and the equal-time anticommutation relation
\be
\left\{ \lb_+(t,x), \lb_+(t,y) \right\} = \del(x - y).
\label{2.14}
\ee
Defining the plane-wave expansion by 
\be
\lb_+(t+x) = \int_0^{\infty} \frac{dk}{2\pi} \lh \ag_k e^{-ik(t+x)} + \ag^*_k e^{ik(t+x)} \rh,
\label{2.15}
\ee
with inverse 
\be
\ag_k = \frac{i}{2k}\, \int_{-\infty}^{\infty} dx\, e^{ik(t+x)} \stackrel{\leftrightarrow}{\der}_t \lb_+, \hs{2}
\ag^*_k = - \frac{i}{2k}\, \int_{-\infty}^{\infty} dx\, e^{-ik(t+x)} \stackrel{\leftrightarrow}{\der}_t \lb_+,
\label{2.16}
\ee
the anti-commutation relations translate to 
\be
\left\{ \ag_k, \ag^*_q \right\} = 2\pi \del (k-q).
\label{2.17}
\ee
After normal-ordering the hamiltonian (\ref{2.13}) becomes 
\be
{:H}_f:\, = \int_0^{\infty} \frac{dk}{2\pi}\, k \ag^*_k \ag_k,
\label{2.18}
\ee
with ground state the fermionic Minkowski vacuum $| 0 \rangle_f$: 
\be
\forall k: \hs{1} \ag_k | 0 \rangle_f = 0.
\label{2.19}
\ee
The theory of a combined massless scalar and spinor possesses supersymmetry.
The combined hamiltonian
\be
H = {:H}_b: +\, {:H}_f:\, = \int_0^{\infty} \frac{dk}{2\pi}\, k \lh a^*_k a_k + \ag^*_k \ag_k \rh,
\label{2.20}
\ee
allows a square root 
\be
Q = \int_0^{\infty} \frac{dk}{2\pi} \sqrt{k} \lh a^*_k \ag_k + \ag^*_k a_k \rh, 
\label{2.21}
\ee
such that  $Q^2 = H$ and $\left[ Q, H \right] = 0$. The supersymmetry transformations of
the mode operators have the standard form
\be
\ba{ll}
\left[ a_k, Q \right] = \sqrt{k}\, \ag_k, & \left[ a^*_k, Q \right] = - \sqrt{k}\, \ag^*_k, \\
 & \\
\left\{ \ag_k, Q \right\} = \sqrt{k}\, a_k, & \left\{ \ag^*_k, Q \right\} = \sqrt{k}\, a^*_k. 
\ea
\label{2.22}
\ee
The common Minkowski vacuum $| 0 \rangle_M = | 0 \rangle_b \otimes | 0 \rangle_f$ 
is supersymmetric:
\be
Q | 0 \rangle_M = 0.
\label{2.23}
\ee
It is straightforward to generalize the contructions to theories of a larger number of 
bosons and fermions. With equal numbers of $N$ bosons and $N$ fermions there 
are in general $N!$ supersymmetries, associating every boson with each of the 
fermions, as at this point the various boson and fermion fields are indistinguishable
and related by an $O(N) \otimes O(N)$ symmetry.
\vs{2}

\nit
{\bf \large 3.\ Massless fields in 2-D Rindler space}
\vs{1}

\nit
Two-dimensional Rindler space is the space-time described by the metric (\ref{1.1}), 
with the conformal factor specified for the left and right wedges by (\ref{1.4}). It is 
therefore obvious, that in Rindler space 
\[
\sqrt{-g}\, g^{\mu\nu} = \eta^{\mu\nu}.
\]
Moreover the scalar field in the Rindler frame is related to the original Minkowski field by 
the co-ordinate transformation (\ref{1.3}): $\vf_R(\tau, \xi) = \vf[t(\tau,\xi), x(\tau,\xi)]$. 
It follows that the action for scalar fields in Rindler space is formally identical to that 
in Minkowski space:
\be
S_{Rb} = \frac{1}{2}\, \int_R d\tau d\xi \left[\lh \der_{\tau} \vf \rh^2 - \lh \der_{\xi} \vf \rh^2 \right].
\label{3.1}
\ee
which holds both in the right and left wedge of Rindler space, independent of the 
different conformal factors $\rg = e^{\pm a\xi}$. Concentrating again on the left-moving 
solutions of the associated Klein-Gordon equation and following the same procedures 
as described for bosons in section 2, we can repesent them in terms of the plane-wave 
expansion 
\be
\vf_{R+}(\tau + \xi) = \int_0^{\infty} \frac{d\kg}{2\pi \sqrt{2\kg}} \lh b_{\kg} e^{-i\kg (\tau + \xi)} 
 + b^*_{\kg} e^{i\kg(\tau + \xi)} \rh, 
\label{3.2}
\ee
where 
\be
b_{\kg} = \frac{i}{\sqrt{2\kg}}\, \int_{-\infty}^{\infty} d\xi\, e^{i\kg(\tau + \xi)} 
 \stackrel{\leftrightarrow}{\der}_{\tau} \vf_{R+}, \hs{1}
b^*_{\kg} = - \frac{i}{\sqrt{2\kg}}\, \int_{-\infty}^{\infty} d\xi\, e^{-i\kg(\tau + \xi)}
\stackrel{\leftrightarrow}{\der}_{\tau} \vf_{R+}.
\label{3.3}
\ee
Free-field dynamics is generated by introducing the hamiltonian 
\be
H_{Rb} = \int_0^{\infty} \frac{d\kg}{4\pi}\, \kg \lh b^*_{\kg} b_{\kg} + b_{\kg} b^*_{\kg} \rh,
\label{3.4}
\ee
together with the commutation relation 
\be
\left[ b_{\kg}, b^*_{\sg} \right] = 2 \pi \del(\kg - \sg).
\label{3.5}
\ee
Standard procedure would replace $H_{Rb}$ by its normal-ordered form, but normal
ordering is now ambiguous as the result is different in terms of the mode operators 
$(b_{\kg}, b^*_{\kg})$ or in terms the original mode operators $(a_k, a^*_k)$. Indeed, 
the Rindler vacuum $|0 \rangle_{Rb}$ defined by 
\be
b_{\kg} | 0 \rangle_{Rb} = 0,
\label{3.6}
\ee
is not the same as the bosonic vacuum $| 0 \rangle_b$ in the original Minkowski space, 
as discussed below.
\vs{1}

\nit
Translating the free fermion theory to Rindler space is somewhat more involved. 
Starting from the Dirac-Majorana action on a general space-time manifold $\cM$
\be
S_{\cM f} = \frac{i}{2}\, \int_{\cM} d^2x\, \sqrt{-g}\, \lb^T C \sDer \lb, \hs{2}
\sDer = \gam^a e_a^{\,\mu} \lh \der_{\mu} - \frac{1}{2}\, \og_{\mu}^{\,ab} \sg_{ab} \rh,
\label{3.7}
\ee
where $e_a^{\,\mu}$ is the inverse 2-bein, $\og_{\mu}^{\,ab}$ the associated spin 
connection and $\sg_{ab}$ the generators of 2-D Lorentz transformations in tangent 
spinor space; as explained in appendix A we find that in a conformally flat space-time 
(\ref{1.1}) the Dirac operator in the Majorana representation takes the form 
\be
\sqrt{-g}\, C \sDer = \sqrt{\rg} \lh \ba{cc} \der_{\tau} - \der_{\xi} & 0 \\ 0 & 
 \der_{\tau} + \der_{\xi}  \ea \rh \sqrt{\rg}.  
\label{3.8}
\ee
In addition, the transformation from Minkowski to Rindler co-ordinates also requires 
a $\tau$-dependent Lorentz transformation of the spinor fields; this is also explained 
in appendix A. The upshot is, that upon redefining the spinor fields in the $R$- and 
$L$-wedge of Rindler space by
\be
\ba{l}
R: \hs{2} \psi_{\pm}(\tau,\xi) = e^{\pm a(\tau \pm \xi)/2} \lb_{\pm}[t(\tau,\xi), x(\tau,\xi)], \\
 \\
L: \hs{2} \psi_{\pm}(\tau,\xi) = e^{\mp a(\tau \pm \xi)/2} \lb_{\pm}[t(\tau,\xi), x(\tau,\xi)],
\ea
\label{3.9}
\ee
the Dirac-Majorana action in a conformally flat space-time like Rindler space reduces to 
\be
S_{Rf} = \frac{i}{2}\, \int_R d\tau d\xi \left[ \psi_+ \lh \der_{\tau} - \der_{\xi} \rh \psi_+
 + \psi_- \lh \der_{\tau} + \der_{\xi} \rh \psi_- \right].
\label{3.10}
\ee
The spinor fields $\psi_{\pm}$ have the same formal properties in Rindler space as the 
spinor fields $\lb_{\pm}$ in Minkowski space; concentrating on the left-moving field we 
introduce the plane-wave expansion 
\be
\psi_+(\tau + \xi) = \int_0^{\infty} \frac{d\kg}{2\pi} \lh \bg_{\kg} e^{-i\kg(\tau + \xi)} 
 + \bg^*_{\kg} e^{i\kg(\tau + \xi)} \rh,
\label{3.11}
\ee
where
\be
\bg_{\kg} = \frac{i}{2\kg}\, \int_{-\infty}^{\infty} dx\, e^{i\kg(\tau + \xi)} 
 \stackrel{\leftrightarrow}{\der}_{\tau} \psi_+, \hs{2}
\bg^*_{\kg} = - \frac{i}{2\kg}\, \int_{-\infty}^{\infty} dx\, e^{-i\kg(\tau + \xi)} 
 \stackrel{\leftrightarrow}{\der}_{\tau} \psi_+.
\label{3.12}
\ee
In terms of mode operators the dynamics is defined by the hamiltonian 
\be
H_{Rf} = \int_0^{\infty} \frac{d\kg}{4\pi}\, \kg \lh \bg^*_{\kg} \bg_{\kg} - \bg_{\kg} \bg^*_{\kg} \rh,
\label{3.13}
\ee
and the anti-commutator 
\be
\left\{ \bg_{\kg}, \bg^*_{\sg} \right\} = 2\pi \del(\kg - \sg).
\label{3.14}
\ee
Again, in the pure-fermion theory we renounce normal ordering in view of 
its ambiguity. However, in the special case of a supersymmetric theory with 
equal numbers of bosons and fermions the normal-ordering contributions of 
bosons and fermions cancel, and therefore normal ordering becomes 
unambiguous. In that case we can take the full hamiltonian to be 
\be
H_R = H_{Rb} + H_{Rf} =  
 \int_0^{\infty} \frac{d\kg}{2\pi}\, \kg \lh b^*_{\kg} b_{\kg} + \bg^*_{\kg} \bg_{\kg} \rh.
\label{3.15}
\ee
Clearly it is unique owing to the existence of the supercharge 
\be
Q_R = \int_0^{\infty} \frac{d\kg}{2\pi} \sqrt{\kg} \lh b^*_{\kg} \bg_{\kg} + \bg^*_{\kg} b_{\kg} \rh, 
\label{3.16}
\ee
with the algebraic properties 
\be
Q_R^2 = H_R, \hs{2} \left[ Q_R, H_R \right] = 0.
\label{3.17}
\ee
\vs{1}

\nit
{\bf \large 4.\ Relating Rindler and Minkowski fields} 
\vs{1}

\nit 
In the Rindler wedges the field theories of section 2 and section 3 are related 
by the co-ordinate transformation (\ref{1.3}); they describe the same physics in 
different frames: with respect to an inertial observer using the Minkowski metric, 
or with respect to an accelerated observer using the conformally flat metrics 
(\ref{1.4}). Therefore also the Hilbert spaces and the observables in the two 
descriptions are related; the relation actually is provided by a Bogoliubov 
transformation between the mode operators 
\ct{unruh:1976, hawking:1975, brout:1995, crispino:2008, wald:1984, fulling:1989}.

For the scalar fields the relation is established starting from (\ref{3.3}) and 
inserting for $\vf_{R+}$ its Minkowski space expression, eq.\ (\ref{2.6}). Choosing
units to fix $a = 1$ in the following and introducing the Minkowski light-cone 
co-ordinate
\be
z = t + x = e^{\tau + \xi},
\label{4.1}
\ee
the Rindler mode operators in the $R$-wedge are expressed in terms of the Minkowski 
ones by
\be
\ba{l}
\dsp{ b_{R\kg} = \frac{1}{2\pi\sqrt{2\kg}}\, \int_0^{\infty} \frac{dk}{\sqrt{2k}} \int_0^{\infty} 
 \frac{dz}{z}\, z^{i\kg} \left[ (\kg + kz) a_k e^{-ikz} + (\kg - kz) a^*_k e^{ikz} \right], }\\
 \\
\dsp{ b^*_{R\kg} = - \frac{1}{2\pi\sqrt{2\kg}}\, \int_0^{\infty} \frac{dk}{\sqrt{2k}} \int_0^{\infty} 
 \frac{dz}{z}\, z^{-i\kg} \left[ (\kg - kz) a_k e^{-ikz} + (\kg + kz) a^*_k e^{ikz} \right]. }
\ea
\label{4.2}
\ee
By appropriate choice of contour the integrals over $z$ can be performed in the complex plane;
this leads to 
\be
\ba{l}
\dsp{ b_{R\kg} = - \frac{i}{2\pi \sqrt{\kg}}\, \Gam(1 + i \kg)\, \int_0^{\infty} dk\, k^{-\frac{1}{2} - i \kg} 
  \lh a_k e^{\pi\kg/2} + a^*_k e^{- \pi\kg/2} \rh, }\\
 \\
\dsp{ b^*_{R\kg} = \frac{i}{2\pi \sqrt{\kg}}\, \Gam(1 - i \kg)\, \int_0^{\infty} dk\, k^{-\frac{1}{2} + i \kg} 
  \lh a_k e^{-\pi\kg/2} + a^*_k e^{\pi\kg/2} \rh. }
\ea
\label{4.3}
\ee
A similar calculation for the $L$-wedge, in which $z = - (t+x) = e^{-(\tau + \xi)}$ gives 
\be
\ba{l}
\dsp{ b_{L\kg} = \frac{i}{2\pi \sqrt{\kg}}\, \Gam(1 - i \kg)\, \int_0^{\infty} dk\, k^{-\frac{1}{2} + i \kg} 
  \lh a_k e^{\pi\kg/2} + a^*_k e^{- \pi\kg/2} \rh, }\\
 \\
\dsp{ b^*_{L\kg} = - \frac{i}{2\pi \sqrt{\kg}}\, \Gam(1+ i \kg)\, \int_0^{\infty} dk\, k^{-\frac{1}{2} - i \kg} 
  \lh a_k e^{-\pi\kg/2} + a^*_k e^{\pi\kg/2} \rh. }
\ea
\label{4.4}
\ee
Applying these operators to the bosonic Minkowski vacuum one establishes that 
\be
\ba{l}
\dsp{ e^{\pi\kg/2} b_{R\kg} | 0 \rangle_b = e^{-\pi\kg/2} b^*_{L\kg} | 0 \rangle_b = 
  - \frac{i}{2\pi \sqrt{\kg}}\, \Gam(1 + i \kg)\, \int_0^{\infty} dk\, k^{-\frac{1}{2} - i \kg} a^*_k | 0 \rangle_b, }\\
 \\
\dsp{ e^{- \pi\kg/2} b^*_{R\kg} | 0 \rangle_b = e^{\pi\kg/2} b_{L\kg} | 0 \rangle_b = 
  \frac{i}{2\pi \sqrt{\kg}}\, \Gam(1 - i \kg)\, \int_0^{\infty} dk\, k^{-\frac{1}{2} + i \kg} a^*_k | 0 \rangle_b. }
\ea
\label{4.5}
\ee
It follows that the Minkowski vacuum contains Rindler excitations, and that annihilation of 
an excitation in the right Rindler wedge has the same effect modulo a factor $e^{-\pi\kg}$ 
as creation of one in the left Rindler wedge, and vice versa. \\

\nit
To relate the excitations of the fermion field in Minkowski and Rindler space we proceed 
along similar lines, but with some slight modifications to keep track of the conformal and 
local Lorentz transformations relating the fields in the different observer frames. The 
starting point is provided by the equations (\ref{3.12}), in which we replace the Rindler 
spinor field $\psi_+$ by its Minkowski counterpart (\ref{3.9}). In the $R$-wedge of 
Rindler space, using the light-cone variable (\ref{4.1}), we get 
\be 
\ba{l}
\dsp{ \bg_{R\kg} = \frac{i}{4\pi\kg} \int_0^{\infty} dk \int_0^{\infty} \frac{dz}{z}\, z^{\frac{1}{2} + i \kg}
 \left[ \lh \frac{1}{2} - i \kg - ikz \rh \ag_k e^{-ikz} \rd }\\
 \\
\dsp{ \hs{14} \ld +\, \lh \frac{1}{2} - i\kg + ikz \rh \ag^*_k e^{ikz} \right], }\\
 \\
\dsp{ \bg^*_{R\kg} = - \frac{i}{4\pi\kg} \int_0^{\infty} dk \int_0^{\infty} \frac{dz}{z}\, z^{\frac{1}{2} - i \kg}
 \left[ \lh \frac{1}{2} + i \kg - ikz \rh \ag_k e^{-ikz} \rd }\\
 \\
\dsp{ \hs{14.5} \ld +\, \lh \frac{1}{2} + i\kg + ikz \rh \ag^*_k e^{ikz} \right]. }
\ea
\label{4.6}
\ee
Performing the integrations over $z$ in the compex plane this yields
\be
\ba{l}
\dsp{ \bg_{R\kg} = e^{-i\pi/4} \Gam \lh \frac{1}{2} + i \kg \rh \int_0^{\infty} \frac{dk}{2\pi}\, k^{-\frac{1}{2} - i\kg}  
  \lh \ag_k e^{\pi\kg/2} + i \ag^*_k e^{-\pi\kg/2} \rh, }\\
  \\
\dsp{ \bg^*_{R\kg} = e^{-i\pi/4} \Gam \lh \frac{1}{2} - i \kg \rh \int_0^{\infty} \frac{dk}{2\pi}\, k^{-\frac{1}{2} + i\kg}  
  \lh \ag_k e^{-\pi\kg/2} + i \ag^*_k e^{\pi\kg/2} \rh. }
\ea
\label{4.7}
\ee
Similarly in the $L$-wedge of Rindler space we get after the appropriate modifications
\be
\ba{l}
\dsp{ \bg_{L\kg} = e^{i\pi/4} \Gam \lh \frac{1}{2} - i \kg \rh \int_0^{\infty} \frac{dk}{2\pi}\, k^{-\frac{1}{2} + i\kg}  
  \lh \ag_k e^{\pi\kg/2} - i \ag^*_k e^{-\pi\kg/2} \rh, }\\
  \\
\dsp{ \bg^*_{L\kg} = e^{i\pi/4} \Gam \lh \frac{1}{2} + i \kg \rh \int_0^{\infty} \frac{dk}{2\pi}\, k^{-\frac{1}{2} - i\kg}  
  \lh \ag_k e^{-\pi\kg/2} - i \ag^*_k e^{\pi\kg/2} \rh. }
\ea
\label{4.8}
\ee
From this we derive the following relations for the action on the fermionic Minkowski vacuum: 
\be
\ba{l}
\dsp{ e^{\pi \kg/2} \bg_{R\kg} | 0 \rangle_f = - i e^{-\pi\kg/2} \bg^*_{L\kg} | 0 \rangle_f = 
 e^{i\pi/4} \Gam \lh \frac{1}{2} + i \kg \rh \int_0^{\infty} \frac{dk}{2\pi}\, k^{-\frac{1}{2} - i\kg} 
 \ag^*_k | 0 \rangle_f, }\\
 \\
\dsp{ e^{-\pi \kg/2} \bg^*_{R\kg} | 0 \rangle_f = i e^{\pi\kg/2} \bg_{L\kg} | 0 \rangle_f = 
 e^{i\pi/4} \Gam \lh \frac{1}{2} - i \kg \rh \int_0^{\infty} \frac{dk}{2\pi}\, k^{-\frac{1}{2} + i\kg} 
 \ag^*_k | 0 \rangle_f. }
\ea
\label{4.9}
\ee
Again, the fermionic Minkowski vacuum contains fermionic Rindler excitations, with 
creation of one in the $R$-wedge being accompanied by annihilation of a similar 
one in the $L$-wedge, and vice versa. 

As concerns correlation functions, in free field theories the only non-trivial correlators 
are two-point functions. In Minkowski space the elementary ones are 
\be
{_b\langle} 0| a_k a^*_q | 0 \rangle_b = {_f\langle} 0| \ag_k \ag^*_q | 0 \rangle_f = 2\pi \del(k-q).
\label{4.10}
\ee
Using the results (\ref{4.5}) and (\ref{4.9}) in a recursive way \ct{crispino:2008} one establishes 
that the correlators of the Rindler fields in the Minkowski vacuum are 
\be
{_b\langle} 0| b_{R\kg} b^*_{R\sg} | 0 \rangle_b = \frac{2\pi \del(\kg - \sg)}{1 - e^{-2\pi \kg}}, \hs{2}
{_f\langle} 0| \bg_{R\kg} \bg^*_{R\sg} | 0 \rangle_f = \frac{2\pi \del(\kg - \sg)}{1 + e^{-2\pi \kg}},
\label{4.11}
\ee
with identical results for the correlators in the $L$-wedge. These relations imply the 
complementary identities 
\be
{_b\langle} 0| b^*_{R\kg} b_{R\sg} | 0 \rangle_b = \frac{2\pi \del(\kg - \sg)}{e^{2\pi \kg} - 1}, \hs{2}
{_f\langle} 0| \bg^*_{R\kg} \bg_{R\sg} | 0 \rangle_f = \frac{2\pi \del(\kg - \sg)}{e^{2\pi \kg} + 1},
\label{4.12}
\ee
stating that the occupation numbers of Rindler excitations in the Minkowski vacuum are of
thermal Bose-Einstein and Fermi-Dirac type, corresponding to a state with temperature 
$T = 1/2\pi$; upon restoring the acceleration parameter by rescaling of the Rindler 
co-ordinates and momenta the temperature of Minkowski space observed in a Rindler 
frame with proper acceleration $a$ is $T = a/2\pi$. Details of the derivation are presented
in appendix B. 
\vs{2}

\nit
{\bf \large 5.\ Supersymmetry}
\vs{1}

\nit
As we have seen the supersymmetric theory has a unique Rindler hamiltonian (\ref{3.15}), 
which is the square of the supercharge (\ref{3.16}). In view of the results (\ref{4.12}) it is 
easily seen to have a divergent Minkowski vacuum expectation value. It follows automatically
that the Rindler supercharge does not annihilate the Minkowski vacuum: Rindler supersymmetry
is broken by finite temperature effects in the Minkowski vacuum. In contrast, Minkowski 
supersymmetry generated by the supercharge (\ref{2.21}) does annihilate the Minkowski 
vacuum as stated by (\ref{2.23}).

One way to deal with this situation it is to regularize the Rindler supercharge using a 
symmetric regulator function $g(\kg,\sg) = g(\sg,\kg)$ and define 
\be
Q_{Rg} = \int_0^{\infty} \frac{d\kg}{2\pi} \int_0^{\infty} \frac{d\sg}{2\pi}\, g(\kg,\sg) 
 \lh b^*_{\kg} \bg_{\sg} + \bg^*_{\sg} b_{\kg} \rh.
\label{5.1}
\ee
A straightforward calculations shows that 
\be
Q^2_{Rg} = \int_0^{\infty} \frac{d\kg}{2\pi} \int_0^{\infty} \frac{d\sg}{2\pi}\, g^2(\kg,\sg) 
 \lh b^*_{\kg} b_{\sg} + \bg^*_{\kg} \bg_{\sg} \rh \equiv H_{Rg}, 
\label{5.2}
\ee
where
\be
g^2(\kg,\sg) = \int_0^{\infty} \frac{d\eta}{2\pi}\, g(\kg,\eta) g(\eta,\sg).
\label{5.3}
\ee
This properly regularizes the Rindler supercharge and hamiltonian if $g$ is taken from 
a class of functions having a limit 
\be
g(\kg,\sg) \rightarrow \sqrt{\kg}\, \del(\kg - \sg), \hs{2} g^2(\kg,\sg) \rightarrow \kg\, \del(\kg - \sg).
\label{5.4}
\ee
With such a regulator one easily finds that 
\be
\ba{l}
{_M\langle} 0| Q^2_{Rg} | 0 \rangle_M = {_M\langle} 0| H_{Rg} | 0 \rangle_M \\
  \\
\dsp{\hs{3} =\, \int_0^{\infty} \frac{d\kg}{2\pi}\, g^2(\kg,\kg) 
  \lh \frac{1}{e^{2\pi\kg} - 1} + \frac{1}{e^{2\pi\kg} + 1} \rh = 
  \int_0^{\infty} \frac{d\kg}{2\pi}\, \frac{g^2(\kg,\kg)}{2 \sinh 2\pi \kg},  }
\ea
\label{5.5}
\ee
which is a positive number for any acceptable regulator $g(\kg,\sg)$.
\vs{2}

\nit
{\bf \large 6.\ Conformal symmetries} 
\vs{1}

\nit
In addition to the hamiltonian and the supercharge, which are conserved for the 
theories in the Minkowski frame, one can construct an infinite set of conformal charges 
annihilating the vacuum state which define a continuum generalization of the Virasoro 
and super-Virasoro algebra.

The charges, which can be decomposed in contributions from bosons and from fermions, 
are labeled by the momentum variable $k$; for $k \geq 0$ they annihilate the Minkowski 
vacuum, being defined for a single boson and fermion field by
\be
\ba{l}
\dsp{ L^b_k = \int_0^{\infty} \frac{dq}{2\pi} \sqrt{q(k+q)}\, a^*_q a_{k+q} + 
  \int_0^k \frac{dq}{4\pi} \sqrt{q(k-q)}\, a_q a_{k-q}, }\\
 \\
\dsp{ L^f_k = \int_0^{\infty} \frac{dq}{2\pi} \lh q + \frac{k}{2} \rh \ag^*_q \ag_{k+q} -
   \int_0^k \frac{dq}{4\pi}\, q\, \ag_q \ag_{k-q}, }\\
\ea
\label{6.1}
\ee
whilst those labeled by a negative momentum $-k$ are the hermitean conjugates 
$L_{-k} = L^*_k$. Introducing the general notation $a_{-k} = a^*_k$, $\ag_{-k} = \ag^*_k$, 
it is possible to write these expressions in a short-hand notation using normal ordering:
\be
L^b_k = \int_{-\infty}^{\infty} \frac{dq}{4\pi} \sqrt{|q(q-k)|} :a_q a_{k-q}:\,, \hs{2}
L^f_k = \int_{-\infty}^{\infty} \frac{dq}{4\pi} \lh - q + \frac{k}{2} \rh :\ag_q \ag_{k-q}:.
\label{6.1.a}
\ee
But apart from the fact that the natural domain of the momentum labels $k,q$ is 
$[0, \infty)$, the normal ordering in Rindler space is ambiguous due to the mixing 
of creation an annhihilation operators by the Bogoliubov tranformations, as pointed 
out before. This is especially relevant as we intend to compute correlation functions 
of Rindler operators in the Minkowski vacuum. Therefore we prefer to write out the 
explicit expansions (\ref{6.1}) rather than using the normal-ordered ones. 

The operators (\ref{6.1}) satisfy the commutation relations of the Virasoro 
algebra with a central charge:
\be
\ba{l}
\dsp{ \left[ L^b_k, L^b_l \right] = (k - l) L^b_{k+l} + \frac{1}{12}\, k^3\, \del(k+l), }\\
 \\
\dsp{ \left[ L^f_k, L^f_l \right] = (k - l) L^f_{k+l} + \frac{1}{24}\, k^3\, \del(k+l). }
\ea
\label{6.2}
\ee
The expressions for the central charges are easily verified by the standard 
procedure of evaluating the vacuum expectation value of the commutator 
\[
\frac{1}{2}\, {_b\langle} 0 | \left[ L^b_k, L^b_l \right] |0 \rangle_b = 
 {_f\langle} 0 | \left[ L^f_k, L^f_l \right] |0 \rangle_f = \frac{1}{24}\, k^3\, \del(k+l).
\]
In supersymmetric Minkowski-frame theories the Virasoro algebra can be extended by 
supercharges $G_k$, defined for a single boson plus fermion field and for $k \geq 0$ by 
\be
G_k = \int_0^{\infty} \frac{dq}{2\pi} \lh \sqrt{q}\, a^*_q \ag_{k+q} + \sqrt{k+q}\, \ag^*_q a_{k+q} \rh 
  + \int_0^k \frac{dq}{2\pi} \sqrt{q}\, a_q \ag_{k-q};
\label{6.3}
\ee
again for negative index $G_{-k} = G^*_k$. In a theory with a single boson and fermion 
the complete set of (anti-)commutation relations with 
\[
L_k = L_k^b + L_k^f, 
\]
is found to have the standard super-Virasoro form
\be
\ba{l}
\dsp{ \left[ L_k, L_l \right] = (k - l)\, L_{k+l} + \frac{1}{8}\, k^3 \del(k+l),}\\
 \\
\dsp{ \left[ L_k, G_l \right] = \lh \frac{k}{2} - l \rh G_{k+l}, }\\
 \\
\dsp{ \left\{ G_k, G_l \right\} = 2 L_{k+l} + \frac{1}{2}\, k^2 \del(k+l). }
\ea
\label{6.4}
\ee
Note that $L_0 = H$ is the hamiltonian (\ref{2.20}), and $G_0 = Q$ is the supercharge 
(\ref{2.21}). They define a closed anomaly-free subalgebra of the super-Virasoro 
algebra (\ref{6.4}). Note also that their vacuum expectations vanish: 
\be
\ba{ll}
\forall k \geq 0: & L_k | 0 \rangle_M = 0 \hs{1} \Rightarrow \hs{1}
 {_M\langle} 0| L_{k} |0\rangle_M =  {_M\langle} 0| L_{-k} |0\rangle_M = 0, \\
 & \\
 & G_k | 0 \rangle_M = 0 \hs{1} \Rightarrow \hs{1}
 {_M\langle} 0| G_{k} |0\rangle_M = {_M\langle} 0| G_{-k} |0\rangle_M = 0.
\ea
\label{6.5a}
\ee
The same construction also works in the Rindler frame, where in the $R$-wedge 
of Rindler space
\be
\ba{l}
\dsp{ L^b_{R\kg} \equiv \int_0^{\infty} \frac{d\sg}{2\pi} \sqrt{\sg(\kg+\sg)}\, b^*_{R\sg} b_{R\,\kg+\sg} 
 + \int_0^{\kg} \frac{d\sg}{4\pi} \sqrt{\sg(\kg-\sg)}\, b_{R\sg} b_{R\,\kg-\sg}, }\\
 \\
\dsp{ L^f_{R\kg} = \int_0^{\infty} \frac{d\sg}{2\pi} \lh \sg + \frac{\kg}{2} \rh \bg^*_{R\sg} \bg_{R\,\kg+\sg} 
 - \int_0^{\kg} \frac{d\sg}{4\pi}\, \sg\, \bg_{R\sg} \bg_{R\,\kg-\sg}, }\\
 \\
\dsp{ G_{R\kg} =  \int_0^{\infty} \frac{d\sg}{2\pi} \lh \sqrt{\sg}\, b^*_{R\sg} \bg_{R\, \kg+\sg} + 
 \sqrt{\kg+\sg}\, \bg^*_{R\sg} b_{R\,\kg+\sg} \rh + \int_0^{\kg} \frac{d\sg}{2\pi} \sqrt{\sg}\, b_{R\sg} 
 \bg_{R\,\kg-\sg}, }
\ea
\label{6.5}
\ee
with analogous definitions in the $L$-wedge. The essential difference with the Minkowski 
charges, is that the Rindler operators (\ref{6.5}) admit a non-vanishing expectation value 
in the Minkowski frame:
\be
\ba{lll}
{_M\langle} 0| L^b_{R\kg} |0\rangle_M & = & \dsp{ \del(\kg)\, \int_0^{\infty} d\sg\, 
 \frac{\sg}{e^{2\pi\sg} - 1} = \frac{1}{24}\, \del(\kg), }\\
 & & \\
{_M\langle} 0| L^f_{R\kg} |0\rangle_M & = & \dsp{ \del(\kg)\, \int_0^{\infty} d\sg\, 
 \frac{\sg}{e^{2\pi\sg} + 1} = \frac{1}{48}\, \del(\kg), }\\
\ea
\label{6.6}
\ee
whilst in the superconformal case
\[
{_M\langle} 0| G_{R\kg} |0\rangle_M =  0. 
\]
Although one could remove these vacuum expectation values by shifting the ground-state 
energy, this would introduce an extra contribution to the central charge, which is generally 
not desirable. The singular nature of the expectation value of the hamiltonian $H = L_0$ in 
practice requires regularization, as we discussed in section 5.
\vs{2}

\nit
{\bf \large 7.\ Ghosts and local conformal symmetry} 
\vs{1}

\nit
As is well-known, the line element in any topologically trivial 2-D space-time can be 
cast in the form (\ref{1.1}) by appropriate co-ordinate transformations. As 2-D gravity 
is conformally invariant, but non-dynamical, one can therefore interpret the boson and 
fermion field theories discussed above as the gauge-fixed version of a gravitational 
theory with local conformal symmetry. This is the key to a consistent quantum-theory 
of strings and superstrings \ct{green:1987}. Local conformal symmetry turns the 
generators of the Virasoro algebra into operators imposing first-class constraints:
\be
\forall k > 0: \hs{1} L_k | phys \rangle = 0.
\label{7.1}
\ee
These conditions are consistent only if the anomaly vanishes, but the anomaly now 
includes a contribution from the Faddeev-Popov ghosts introduced by quantization 
of the gravitational background as well \ct{siegel:1984}. 

In the sector of left-moving fields the dynamics of the anti-commuting gravity ghosts 
$(C,D)$ follows from the action
\be
S_g = i\, \int_M d^2 x\, D \lh \der_t - \der_x \rh C.
\label{7.2}
\ee
Following standard procedures one derives the hamiltonian 
\be
H_g = i\, \int_{-\infty}^{\infty} dx\, D \der_x C, 
\label{7.3}
\ee
supplemented by the equal-time anti-commutation relations 
\be
\left\{ C(t,x), D(t,y) \right\} = \del(x - y).
\label{7.4}
\ee
From the mode expansions 
\be
\ba{l}
\dsp{ C(t,x) = \int_0^{\infty} \frac{dk}{2\pi} \lh c_k e^{-ik(t+x)} + c^*_k e^{ik(t+x)} \rh, }\\
 \\
\dsp{ D(t,x) = \int_0^{\infty} \frac{dk}{2\pi} \lh d_k e^{-ik(t+x)} + d^*_k e^{ik(t+x)} \rh, }
\ea
\label{7.5}
\ee
we derive the anti-commutators of the mode operators:
\be
\left\{ c_k, d^*_q \right\} = 2\pi \del(k-q), \hs{2} \left\{ d_k, c^*_q \right\} = 2\pi \del(k - q).
\label{7.6}
\ee
The corresponding Virasoro generators for $k \geq 0$ are 
\be
L^g_k = \int_0^{\infty} \frac{dq}{2\pi} \left[ \lh q - k \rh c^*_q d_{k+q} + 
 \lh q + 2k \rh d^*_q c_{k+q} \right] - \int_0^k \frac{dq}{2\pi}\, \lh q + k \rh c_q d_{k-q},
\label{7.7}
\ee
with $L^g_{-k} = L^{g\, *}_k$. The ghost Virasoro algebra then becomes 
\be
\left[ L^g_k, L^g_l \right] = (k-l)\, L^g_{k+l} - \frac{13}{6}\, k^3 \del(k+l).
\label{7.8}
\ee
For a theory with $N_b$ massless scalars and $N_f$ massless chiral Majorana 
fermions the full first-class contraints then are 
\be
\forall k > 0: \hs{1} L_k | phys \rangle = 0, \hs{2} 
 L_k = \sum_{i=1}^{N_b} L^{bi}_k + \sum_{j =1}^{N_f} L^{fj}_k + L^g_k,
\label{7.9}
\ee
with the algebra 
\be 
\left[ L_k, L_l \right] = (k-l)\, L_{k+l} + \frac{(2N_b + N_f - 52)}{24}\, k^3 \del(k+l).
\label{7.10}
\ee
Thus the central charge vanishes provided 
\be
2N_b + N_f = 52.
\label{7.11}
\ee
The analysis in the right-moving sector proceeds entirely analogously. 

In the case of a supersymmetric theory of massless scalars and femions the 
local conformal symmetry can be extended to local superconformal symmetry. 
Such a theory is effectively a gauge-fixed 2-D supergravity theory and requires 
an additional set of commuting supersymmetry ghosts $(S,U)$ with action 
\be
S_{sg} = \int_M d^2 x\, U \lh \der_t - \der_x \rh S. 
\label{7.12}
\ee
The associated hamiltonian and equal-time commutation relations are 
\be
H_{sg} = \int_{-\infty}^{\infty} dx\, U \der_x S, \hs{2} \left[ S(t,x), U(t,y) \right] = i \del(x - y).
\label{7.13}
\ee
In this case we take the plane-wave expansions to be 
\be
\ba{l}
\dsp{ S(t,x) = \int_0^{\infty} \frac{dk}{2\pi} \lh s_k e^{-ik(t+x)} + s^*_k e^{ik(t+x)} \rh, }\\
 \\
\dsp{ U(t,x) = -i\, \int_0^{\infty} \frac{dk}{2\pi} \lh u_k e^{-ik(t+x)} - u^*_k e^{ik(t+x)} \rh, } 
\ea
\label{7.14}
\ee
which results in mode commutation relations
\be
\left[ s_k, u^*_q \right] = 2\pi\, \del(k-q), \hs{2} \left[ u_k, s^*_q \right] = 2\pi\, \del(k-q).
\label{7.15}
\ee
The super-Virasoro operators of the full set of ghosts $(C,D)$ and superghosts $(S,U)$ 
then is defined by (\ref{7.7}) and for $k \geq 0$:
\be
\ba{l}
\dsp{ L^{sg}_k = \int_0^{\infty} \frac{dq}{2\pi} \left[ \lh q - \frac{k}{2} \rh s^*_q u_{k+q} 
 + \lh q + \frac{3k}{2} \rh u^*_q s_{k+q} \right] }\\
 \\
\dsp{ \hs{3} -\, \int_0^k \frac{dq}{2\pi} \lh q + \frac{k}{2} \rh s_q u_{k-q}, }\\
 \\
\dsp{ G^{sg}_k = \int_0^{\infty} \frac{dq}{2\pi} \left[ s^*_q b_{k+q} + b^*_q s_{k+q} 
 + (q + 3 k) u^*_q c_{k+q} + (q-2k) c^*_q u_{k+q} \right] }\\
 \\
\dsp{ \hs{3} +\, \int_0^k \frac{dq}{2\pi} \left[ s_q b_{k-q} - \lh q + 2 k \rh c_q u_{k-q} \right], }
\ea
\label{7.16}
\ee
with $L^{sg}_{-k} = L^{sg\,*}_k$ and $G^{sg}_{-k} = G^{sg\,*}_k$. The complete 
super-Virasoro algebra of the ghosts is given by (\ref{7.8}) and
\be
\ba{l}
\dsp{ \left[ L^{sg}_k, L^{sg}_{l} \right] = (k-l)\, L^{sg}_{k+l} + \frac{11}{12}\, k^3 \del(k+l), }\\
 \\
\dsp{ \left\{ G^{sg}_k, G^{sg}_l \right\} = 2 \lh L^g_{k+l} + L^{sg}_{k+l} \rh - 5 k^2 \del(k+l). }
\ea
\label{7.17}
\ee
With these results a 2-D supersymmetric theory of $N$ massless scalars and fermions has 
local superconformal symmetry if 
\be
\forall k > 0: \hs{1} L_k | phys \rangle = G_k | phys \rangle = 0,
\label{7.18}
\ee
where 
\be
L_k = \sum_{i=1}^N \lh L^{bi}_k + L^{fi}_k \rh + L^g_k + L^{sg}_k, \hs{2}
G_k = \sum_{i = 1}^N G^i_k + G^{sg}_k,
\label{7.19}
\ee
with the algebra 
\be
\ba{l}
\dsp{ \left[ L_k, L_l \right] = (k - l)\, L_{k+l} + \frac{(N-10)}{8}\, k^3 \del(k+l),}\\
 \\
\dsp{ \left[ L_k, G_l \right] = \lh \frac{k}{2} - l \rh G_{k+l}, }\\
 \\
\dsp{ \left\{ G_k, G_l \right\} = 2 L_{k+l} + \frac{(N-10)}{2}\, k^2 \del(k+l). }
\ea
\label{7.20}
\ee
Thus all superconformal anomalies vanish for $N = 10$. Of course all results 
derived here are in agreement with corresponding string and superstring theories 
in the limit of infinite strings.
\vs{2}

\nit
{\bf \large 8.\ Conformal and superconformal ghosts in Rindler space}
\vs{1}

\nit
The local conformal and superconformal symmetries arising in a gauge-fixed 
2-D gravity or supergravity theory with scalar and spinor matter can be 
extended to the Rindler frame. It requires solving the ghost and superghost
field equations in Rindler space, and finding the appropriate Bogoliubov 
tranformations.

Consider a theory in an arbitrary 2-D space-time with reparametrization and 
locally Lorentz-invariant action 
\be
S[F,G] = \int d^2x\, e\, F^{a_1... a_{n+1}} e_{a_{n+1}}^{\;\mu} D_{\mu} G_{a_1...a_n}.
\label{8.1}
\ee
Here $(F,G)$ is a system of commuting or anti-commuting local Lorentz tensors 
or (chiral) spinor-tensors of rank $(n, n+1)$, $e_a^{\;\mu}$ is the inverse 2-bein 
field as explained in appendix A; $D_{\mu}$ is the Lorentz-covariant derivative 
with spin connection $\og_{\mu\;b}^{\;\,a}$ defined in (\ref{a.7}) and 
$e = \det e_{\mu}^{\;a} = \sqrt{-g}$. Consider only purely left- or right-handed 
components associated with purely self-dual or anti-self-dual tensors: 
\be
G_{a_1...a_n} = \pm\, \ve_{a_1b} G^b_{\;\,a_2 ... a_n} = ... 
  = \pm\, \ve_{a_nb} G_{a_1 ... a_{n-1}}^{\hs{2.5} b},
\label{8.2}
\ee
and similarly for $F^{a_1...a_{n+1}}$. Fixing the Minkowski gauge, the action 
(\ref{8.1}) can be decomposed into actions of the type (\ref{7.2}) or (\ref{7.12}) 
for single pairs of components $(F_M^{(n+1)},G_M^{(n)})$ with gauge-fixed 
action (considering left-movers for definiteness)
\be
S^{(n)}_M = \int_M d^2x\, F_M^{(n+1)} \lh \der_t - \der_x \rh G_M^{(n)}.
\label{8.3}
\ee
When transforming the action from the Minkowski to the Rindler frame by the 
transformations (\ref{1.3}) with $a = 1$ and following the same arguments as described 
for spinor components in appendix A, we get a Rindler-frame action in the $R$-wedge 
of Rindler space
\be
S^{(n)}_R = \int_R d\tau d\xi \left[ F^{(n+1)}_R \lh \der_{\tau} - \der_{\xi} \rh G^{(n)}_R \right],
\label{8.4}
\ee
where the field components have been transformed from the Minkowski to the 
Rindler frame by an extension of (\ref{3.9}):
\be
\ba{l}
\dsp{ F^{(n+1)}_R(\tau,\xi) = e^{(n+1)(\tau + \xi)} F^{(n+1)}_M[t(\tau,\xi), x(\tau,\xi)], }\\
 \\
\dsp{ G^{\,n}_R(\tau,\xi) = e^{-n(\tau + \xi)} G^{\,n}_M[t(\tau,\xi), x(\tau,\xi)]. }
\ea
\label{8.5}
\ee
Actually $n$ can be integer or half-integer, depending on whether $(F,G)$ are tensors 
or spinor-tensors. The aim is now to connect the Rindler plane-wave expansions in 
the $R$-wedge
\be
\ba{l}
\dsp{ F_R^{(n+1)} = \int_0^{\infty} \frac{d\kg}{2\pi} \lh f_{R\kg} e^{- i \kg (\tau + \xi)} 
 + f^*_{R\kg} e^{i\kg (\tau + \xi)} \rh }\\
 \\
\dsp{ G^{(n)}_R = \int_0^{\infty} \frac{d\kg}{2\pi} \lh g_{R\kg} e^{- i \kg (\tau + \xi)} 
 + g^*_{R\kg} e^{i\kg (\tau + \xi)} \rh, }
\ea
\label{8.6n}
\ee
with the Minkowski ones
\be
\ba{l}
\dsp{ F_M^{(n+1)} = \int_0^{\infty} \frac{dk}{2\pi} \lh f_{k} e^{- i k (t+x)} 
 + f^*_{\kg} e^{i\kg (t + x)} \rh }\\
 \\
\dsp{ G^{(n)}_M = \int_0^{\infty} \frac{dk}{2\pi} \lh g_{k} e^{- k(t+x)} 
 + g^*_{\kg} e^{ik (t + x)} \rh. }
\ea
\label{8.7n}
\ee
Inverting the equations (\ref{8.6n}) using the light-cone variable $z$ defined in (\ref{4.1}):
\be
\ba{lll}
f_{R\kg} & = & \dsp{ \frac{i}{2\kg}\, \int_{-\infty}^{\infty} d\xi\, e^{i\kg(\tau + \xi)} 
 \stackrel{\leftrightarrow}{\der}_{\tau} F^{(n+1)}_R(\tau,\xi) }\\
 & & \\
 & = & \dsp{  \frac{i}{2\kg}\, \int_0^{\infty} \frac{dz}{z}\, z^{-n-1+i\kg} \lh -i \kg + n + 1 + z \der_z \rh   
 F^{(n+1)}_M(z). }\\
 & & \\
g_{R\kg} & = & \dsp{ \frac{i}{2\kg}\, \int_{-\infty}^{\infty} d\xi\, e^{i\kg(\tau + \xi)} 
 \stackrel{\leftrightarrow}{\der}_{\tau} G^{(n)}_R(\tau,\xi) }\\
 & & \\
 & = & \dsp{  \frac{i}{2\kg}\, \int_0^{\infty} \frac{dz}{z}\, z^{n+i\kg} \lh -i \kg - n + z \der_z \rh 
 G^{(n)}_M(z), }
\ea
\label{8.7}
\ee
Substitutions of the Minkowski-frame plane-wave solutions yields 
\be
\ba{lll}
f_{R\kg} & = & \dsp{ i^{(n+1)} \Gam(n + 1 + i\kg) \int_0^{\infty} \frac{dk}{2\pi}\, k^{-n-1- i\kg} 
 \lh (-1)^{n+1} e^{\pi\kg/2} f_k + e^{-\pi\kg/2} f^*_k \rh, }\\
 & & \\
g_{R\kg} & = & \dsp{ i^{n} \Gam(-n + i\kg) \int_0^{\infty} \frac{dk}{2\pi}\, k^{n - i\kg} 
 \lh e^{\pi\kg/2} g_k + (-1)^n e^{-\pi\kg/2} g^*_k \rh, } 
\ea
\label{8.8}
\ee
with conjugates
\be
\ba{lll}
f^*_{R\kg} & = & \dsp{ i^{(n+1)} \Gam(n + 1 - i\kg) \int_0^{\infty} \frac{dk}{2\pi}\, k^{-n-1+ i\kg} 
 \lh (-1)^{n+1} e^{-\pi\kg/2} f_k + e^{\pi\kg/2} f^*_k \rh, }\\
 & & \\
g^*_{R\kg} & = & \dsp{ i^{n} \Gam(-n - i\kg) \int_0^{\infty} \frac{dk}{2\pi}\, k^{n + i\kg} 
 \lh e^{-\pi\kg/2} g_k + (-1)^n e^{\pi\kg/2} g^*_k \rh. } 
\ea
\label{8.9}
\ee
In the $L$-wedge of Rindler space one finds by analogous calculations for the left-moving 
fields
\be
\ba{lll}
f_{L\kg} & = & \dsp{ i^{(n+1)} \Gam(n +1 - i\kg) \int_0^{\infty} \frac{dk}{2\pi}\, k^{-n-1+ i\kg} 
 \lh e^{\pi\kg/2} f_k + (-1)^{n+1} e^{-\pi\kg/2} f^*_k \rh, }\\
 & & \\
g_{L\kg} & = & \dsp{ i^{n} \Gam(-n - i\kg) \int_0^{\infty} \frac{dk}{2\pi}\, k^{n + i\kg} 
 \lh (-1)^n e^{\pi\kg/2} g_k + e^{-\pi\kg/2} g^*_k \rh. } 
\ea
\label{8.10}
\ee
and conjugates
\be
\ba{lll}
f^*_{L\kg} & = & \dsp{ i^{(n+1)} \Gam(n +1 + i\kg) \int_0^{\infty} \frac{dk}{2\pi}\, k^{-n - 1- i\kg} 
 \lh e^{-\pi\kg/2} f_k + (-1)^{n+1} e^{\pi\kg/2} f^*_k \rh, }\\
 & & \\
g^*_{L\kg} & = & \dsp{ i^{n} \Gam(-n + i\kg) \int_0^{\infty} \frac{dk}{2\pi}\, k^{n - i\kg} 
 \lh (-1)^n e^{-\pi\kg/2} g_k + e^{\pi\kg/2} g^*_k \rh. } 
\ea
\label{8.11}
\ee
Upon application to the Minkowski vacuum it follows that 
\be
\ba{ll}
e^{\pi\kg/2} f_{R\kg} | 0 \rangle_M = (-1)^{n+1} e^{-\pi\kg/2} f^*_{L\kg} | 0 \rangle_M, &
e^{-\pi\kg/2} f^*_{R\kg} | 0 \rangle_M = (-1)^{n+1} e^{\pi\kg/2} f_{L\kg} | 0 \rangle_M, \\
 & \\ 
(-1)^n e^{\pi\kg/2} g_{R\kg} | 0 \rangle_M = e^{-\pi\kg/2} g^*_{L\kg} | 0 \rangle_M, & 
(-1)^n e^{-\pi\kg/2} g^*_{R\kg} | 0 \rangle_M = e^{\pi\kg/2} g_{L\kg} | 0 \rangle_M. 
\ea
\label{8.12}
\ee
Applying these results to the ghost system (\ref{7.2}), which fits the system with $n= 1$
\ct{green:1987,siegel:1984,petcher:1987},  and obeying the anti-commutation rules
\be
\left\{ c_{R\kg}, d^*_{R\sg} \right\} = \left\{ c_{L\kg}, d^*_{L\sg} \right\} = 
\left\{ d_{R\kg}, c^*_{R\sg} \right\} = \left\{ d_{L\kg}, c^*_{L\sg} \right\} = 2 \pi \del(\kg - \sg),
\label{8.13}
\ee
we can compute the two-point correlations of the Rindler ghosts in the Minkowski vacuum:
\be
\ba{l}
\dsp{ {_M\langle} 0 | b_{R\sg} c^*_{R\kg} | 0 \rangle_M = 
 {_M\langle} 0 | c_{R\sg} b^*_{R\kg} | 0 \rangle_M = \frac{2\pi \del(\kg - \sg)}{1 - e^{-2\pi \kg}}, }\\
 \\
\dsp{ {_M\langle} 0 | c^*_{R\kg} b_{R\sg} | 0 \rangle_M = 
 {_M\langle} 0 | b^*_{R\kg} c_{R\sg} | 0 \rangle_M = - \frac{2\pi \del(\kg - \sg)}{e^{2\pi \kg} - 1}. }
\ea
\label{8.14}
\ee
Note that these results are independent of the value of $n$, but do reflect the 
ghost statistics leading to the minus sign in the last expression.

Finally we turn to the commuting superghosts $(S,U)$. The analysis proceeds parallel 
to the ghost system $(C,D)$, except that the appropriate value Lorentz and conformal
weight is $n = 1/2$ \ct{green:1987,vholten:1987} and the anti-commutation rules are 
replaced by commutation rules. The relevant results are 
\be
\ba{ll}
e^{\pi\kg/2} u_{R\kg} | 0 \rangle_M = (-1)^{n+1} e^{-\pi\kg/2} u^*_{L\kg} | 0 \rangle_M, &
e^{-\pi\kg/2} u^*_{R\kg} | 0 \rangle_M = (-1)^{n+1} e^{\pi\kg/2} u_{L\kg} | 0 \rangle_M, \\
 & \\ 
(-1)^n e^{\pi\kg/2} s_{R\kg} | 0 \rangle_M = e^{-\pi\kg/2} s^*_{L\kg} | 0 \rangle_M, & 
(-1)^n e^{-\pi\kg/2} s^*_{R\kg} | 0 \rangle_M = e^{\pi\kg/2} s_{L\kg} | 0 \rangle_M, 
\ea
\label{8.15}
\ee
supplemented by the commutation rules
\be
\left[ s_{R\kg}, u^*_{R\sg} \right] = \left[ s_{L\kg}, u^*_{L\sg} \right] = 
\left[ u_{R\kg}, s^*_{R\sg} \right] = \left[ u_{L\kg}, s^*_{L\sg} \right] = 2\pi \del(\kg-\sg).
\label{8.16}
\ee
From these equations one derives the Minkowski two-point functions for the superghosts:
\be
\ba{l}
\dsp{ {_M\langle} 0 | u_{R\sg} s^*_{R\kg} | 0 \rangle_M = 
 {_M\langle} 0 | s_{R\sg} u^*_{R\kg} | 0 \rangle_M = \frac{2\pi \del(\kg - \sg)}{1 + e^{-2\pi \kg}}, }\\
 \\
\dsp{ {_M\langle} 0 | s^*_{R\kg} u_{R\sg} | 0 \rangle_M = 
 {_M\langle} 0 | u^*_{R\kg} s_{R\sg} | 0 \rangle_M = - \frac{2\pi \del(\kg - \sg)}{e^{2\pi \kg} + 1}. }
\ea
\label{8.17}
\ee

\nit
{\bf \large 9.\ Local conformal invariance in Rindler space}
\vs{1}

\nit
With the results (\ref{8.14}) and (\ref{8.17}) we can now compute the Minkowski 
expectation values of the (super-)Virasoro operators including the ghost contributions 
required for local conformal and superconformal invariance. First, the results (\ref{6.5}) 
are supplemented for $\kg  \geq 0$ by
\be
\ba{l}
\dsp{ L^g_{R\kg} = \int_0^{\infty} \frac{d\sg}{2\pi} \left[ \lh \sg - \kg \rh c^*_{\sg} d_{\kg+\sg} + 
 \lh \sg + 2\kg \rh d^*_{\sg} c_{\kg+\sg} \right] - \int_0^{\kg} \frac{d\sg}{2\pi}\, \lh \sg + \kg \rh 
 c_{\sg} d_{\kg-\sg}, }\\
 \\
\dsp{ L^{sg}_{R\kg} = \int_0^{\infty} \frac{d\sg}{2\pi} \left[ \lh \sg - \frac{\kg}{2} \rh 
 s^*_{\sg} u_{\kg+\sg} + \lh \sg + \frac{3\kg}{2} \rh u^*_{\sg} s_{\kg+\sg} \right] }\\
 \\
\dsp{ \hs{3} -\, \int_0^{\kg} \frac{d\sg}{2\pi} \lh \sg + \frac{\kg}{2} \rh s_{\sg} u_{\kg-\sg}, }\\
 \\
\dsp{ G^{sg}_{R\kg} = \int_0^{\infty} \frac{d\sg}{2\pi} \left[ s^*_{\sg} b_{\kg+\sg} + b^*_{\sg} s_{\kg+\sg} 
 + (\sg + 3 \kg) u^*_{\sg} c_{\kg+\sg} + (\sg-2\kg) c^*_{\sg} u_{\kg+\sg} \right] }\\
 \\
\dsp{ \hs{3} +\, \int_0^{\kg} \frac{d\sg}{2\pi} \left[ s_{\sg} b_{\kg-\sg} - \lh \sg + 2 \kg \rh 
 c_{\sg} u_{\kg-\sg} \right], }
\ea
\label{9.1}
\ee
with $(L^g_R)_{-\kg} = L^{g\,*}_{R\kg}$, $(L^{sg}_R)_{-\kg} = L^{sg\,*}_{R\kg}$ and 
$(G^{sg}_R)_{-\kg} = G^{sg\,*}_{R\kg}$.
Together with the operators (\ref{6.5}) for each scalar and spinor field they define 
the same algebra (\ref{7.20}) in Rindler space. Thus the first-class constraints 
requiring the cancellation of the anomalies are also satisfied for $2N_b + N_f = 52$
bosons and fermions coupled to 2-D conformal gravity and $N_b = N_f = 10$ 
bosons and fermions coupled to 2-D conformal supergravity.

As in the case of global conformal invariance, the Minkowski expectation values
of the conformal charges are modified due to finite temperature effects. With the 
help of equations (\ref{8.15}) and (\ref{8.17}) the results (\ref{6.6}) are now extended 
by 
\be
{_M\langle} 0 | L^g_{R\kg} | 0 \rangle_M = - \frac{1}{12}\, \del(\kg), \hs{2}
{_M\langle} 0 | L^{sg}_{R\kg} | 0 \rangle_M = - \frac{1}{24}\, \del(\kg).
\label{9.2}
\ee
For the full Virasoro charges, simultaneously requiring cancellation of the anomalies:
\be
\ba{l}
{_M\langle} 0 | \lh N_b L^b_{R\kg} + N_f L^f_{R\kg} + L^g_{R\kg} \rh | 0 \rangle_M = \del(\kg), \\
 \\
\dsp{ {_M\langle} 0 | \lh N_b L^b_{R\kg} + N_f L^f_{R\kg} + L^g_{R\kg} + 
 L^{sg}_{R\kg} \rh | 0 \rangle_M = \frac{1}{2}\, \del(\kg). }
\ea
\label{9.3}
\ee
These expectation values, restricted to $\kg = 0$, do not contradict the first-class 
constraints, at the same time showing that the full hamiltonian in Rindler space 
has a non-vanishing Minkowski expectation value as a result of finite-temperature 
correlations (\ref{8.14}), (\ref{8.17}). As discussed before, a regularization procedure 
is required to deal with the $\del$-function singularity of the expectation values in 
applications. 
\vs{2}

\np
\nit
{\bf \large 10.\ Conclusions} 
\vs{1}

\nit
Two-dimensional conformal field theories can be formulated in an inertial frame
with a Minkowski metric, or in a uniformly accelerated frame with a local conformal 
Rindler metric defined by $\rg = e^{\pm a\xi}$; here the sign differentiates between 
the right and left wedge of Rindler space. Thus in each wedge one obtains two different 
descriptions of the same theory. The switch between the frames not only changes the 
co-ordinates, but Fock-space quantization can proceed parallel in both frames, provided 
an appropriate conformal transformation is applied to the field components as well. 
The Fock spaces in the two frames are related by Bogoliubov transformations of the 
mode operators, effectively giving rise to finite-temperature correlations of the Rindler 
modes for states build on the Minkowski vacuum, which is the true ground state  
of the theory. The temperature is proportional to the acceleration, an effect well-known 
from the original work of Unruh, Davies, Wald, Fulling and others for accelerated 
frames in Minkowski space, and the work of Hawking in the context of black hole 
space-times. 

The switch to the Rindler frame not only creates an effective finite-temperature 
description, but changes the construction and action of symmetry operators like 
conformal and supersymmetry charges on the ground state. The naive conformal 
and supersymmetries in the Rindler frame are broken by finite-temperature effects 
in the Minkowski vacuum, but the underlying Minkowski symmetries are still present 
even though hidden by the Bogoliubov transformations. 

An additional, different issue is the breaking of conformal and supersymmetries 
by anomalies, extensively studied in the context of string theory. In the field-theory
view taken here they break the local conformal and supersymmetry of arising from 
coupling conformal matter to conformally invariant gravity or supergravity. Thus 
consistent coupling to gravity or supergravity is possible only in the case of specific 
critical theories with a strongly limited spectrum of boson and fermion matter fields, 
reflecting the critical dimensions of quantum string theories. This is also relevant 
as the Minkowski-frame and Rindler-frame formulations of massless QFT's
are related by a conformal transformation; if this transformation is jeopardized by 
the conformal anomaly in theories with a non-critical spectrum it may require 
compensating Wess-Zumino type dynamics  \ct{wess:1971}, e.g.\ involving a 
Liouville field. 
\vs{4}

\nit
{\bf Acknowledgement} \\
I thank Bert Schellekens for a preliminary reading of the manuscript.

\np
\nit
{\bf \large Appendix A: 2-D spinor conventions}
\vs{2}

\nit
In this appendix and in the main body of the paper we denote by greek indices $\mu,\nu,...$ 
vector en tensor components in an arbitrary space-time manifold with metric $g_{\mu\nu}$; 
latin indices $a,b, ...$ refer to vector and tensor components in a tangent Minkowski space
with mertric $\eta_{ab}$. The Majorana representation of the Dirac algebra in 2-dimensional 
Minkowski space is defined in terms of the $2 \times 2$ Pauli matrices as follows: 
\be
\gam^0 = i \sg_2, \hs{2} \gam^1 = \sg_1, \hs{2} \gam_3 = \gam^0 \gam^1 = \sg_3.
\label{a.1}
\ee
In 2-dimensional Minkowski space a Dirac spinor is a 2-component object transforming 
under Lorentz transformations with parameter $\og^{ab} = \og \ve^{ab}$ as 
\be
\lb' = e^{\og^{ab} \sg_{ab}/2} \lb = e^{\og \gam_3/2} \lb, \hs{2} 
\sg_{ab} = \frac{1}{4} \left[ \gam_a, \gam_b \right] = - \frac{1}{2}\, \ve_{ab} \gam_3.
\label{a.2}
\ee
Dirac-conjugate spinors are defined by $\bar{\lb} = \lb^{\dagger} \gam_0$. The 
charge-conjugation operator is $C = - C^T = \gam_0$ with the defining property 
\be
C\gam^a C^{-1} = - \gam^{a\, T}.
\label{a.3}
\ee
A self-conjugate Majorana spinor $\lb = \lb^c = C \bar{\lb}^T = \lb^*$ is therefore 
a spinor with real components:
\be
\lb = \lb^c = \left[ \ba{c} \lb_+ \\ \lb_- \ea \right], \hs{2} \lb^*_{\pm} = \lb_{\pm}.
\label{a.4}
\ee
The subscripts $\pm$ define the eigenvalues under the chiral operator $\gam_3$.
Evidently
\be
\bar{\lb} = \lb^T C.
\label{a.5}
\ee
In a general 2-D space-time is the spinors are defined as representations of the 
local Lorentz group, transforming like (\ref{a.2}) with a space-time dependent 
parameter $\og(x)$. At the same time a spinor transforms like a scalar under 
diffeomorphisms: $\lb'(x') = \lb(x)$. To construct the Dirac-operator one uses 
the 2-bein fields $e_{\mu}^a(x)$ such that the metric is factorized as 
\be
g_{\mu\nu} = e_{\mu}^{\,a} e_{\nu}^{\,b}\, \eta_{ab}.
\label{a.6}
\ee
In terms of the 2-bein fields and their inverse components $e_a^{\,\mu}$ the 
spin connection, acting as the gauge field for local Lorentz transformations, 
is defined by an extension of the metric postulate
\be
\nb_{\mu} e_{\nu}^{\,a} = \der_{\mu} e_{\nu}^{\,a} - \Gam_{\mu\nu}^{\;\;\;\lb} e_{\lb}^{\,a} 
 = \og_{\mu\;\,b}^{\;\,a} e_{\nu}^b,
\label{a.7}
\ee
implying that  
\be
\ba{lll}
\og_{\mu}^{\;\,ab} = - \og_{\mu}^{\;ba} & = & \dsp{ \frac{1}{2} \left[ e^{\nu a} 
 \lh \der_{\nu} e_{\mu}^{\; b} - \der_{\mu} e_{\nu}^{\; b} \rh 
 - e^{\nu b} \lh \der_{\nu} e_{\mu}^{\; a} - \der_{\mu} e_{\nu}^{\; a} \rh \rd }\\
 & & \\
 & & \dsp{ \hs{1.2} \ld +\, e^{\lb a} e^{\nu b} e_{\mu c} 
 \lh \der_{\lb} e_{\nu}^{\; c} - \der_{\nu} e_{\lb}^{\; c} \rh \right]. }
\ea
\label{a.8}
\ee
In terms of these objects the full space-time and Lorentz covariant derivative of a spinor field is 
\be
D_{\mu} \lb = \lh \der_{\mu} - \frac{1}{2}\, \og_{\mu}^{\,ab} \sg_{ab} \rh \lb,
\label{a.9}
\ee
the Dirac operator being defined by $\sDer = \gam^a e_a^{\,\mu} D_{\mu}$.

Next turn to the special class of conformally flat metrics (\ref{1.1}). For these metrics (in
a hybrid notation)
\be
e_{\mu}^{\,a} = \rg\, \del_{\mu}^a, \hs{2} e_a^{\,\mu} = \frac{1}{\rg}\, \del_a^{\mu}.
\label{a.10}
\ee
It follows that the components of the spin connection are
\be
\og_{\mu}^{\; ab} = \lh \del_{\mu}^b \eta^{a\nu} - \del_{\mu}^a \eta^{b\nu} \rh \der_{\nu} \ln \rg.
\label{a.11}
\ee
Specifically in 2-D space-time with conformally flat co-ordinates $(\tau, \xi)$:
\be
\og_{\tau}^{\;01} = - \og_{\tau}^{\;10} = - \der_{\xi} \ln \rg, \hs{1}
\og_{\xi}^{\;01} = - \og_{\xi}^{\;10} = - \der_{\tau} \ln \rg.
\label{a.12}
\ee
Combining this result with (\ref{a.9}) and (\ref{a.2}), while noting that 
\[
\sqrt{-g} = \det e_{\mu}^a = \rg^2, 
\]
one derives the result (\ref{3.8}):
\[
\sqrt{-g}\, C \sDer = \sqrt{\rg} \lh \ba{cc} \der_{\tau} - \der_{\xi} & 0 \\ 0 & 
 \der_{\tau} + \der_{\xi}  \ea \rh \sqrt{\rg}.
\]
But this is not yet the final result; starting form the Minkowski 2-bein 
$e_{M \mu}^{\;\;\;\;a} = \del_{\mu}^a$, the Rindler 2-bein is obtained by
\be
e_{R \mu}^{\;\;\; a} = \dd{x_M^{\lb}}{x_R^{\mu}} e_{M \lb}^{\;\;\;\;b}\, \Og_b^{\;a},
\label{a.13}
\ee
where in the $R$-wedge of Rindler space
\[
\dd{x_M^{\lb}}{x_R^{\mu}} = e^{a\xi} \lh \ba{cc} \cosh a\tau & \sinh a \tau \\
                                                                                \sinh a\tau & \cosh a\tau \ea \rh
\]
is the Jacobian of the transformation from Minkowski to Rindler co-ordinates, and 
$\Og_b^{\;a}$ is a local Lorentz transformation. Clearly the Jacobian is not 
diagonal, but this can be restored by taking the local Lorentz transformation 
\[
\Og = \lh \ba{cc} \cosh a\tau & - \sinh a \tau \\
                          - \sinh a\tau & \cosh a\tau \ea \rh.
\]
This puts the 2-bein back to the diagonal form $e_{R \mu}^{\;\;\;a} = e^{a\xi} \del_{\mu}^a$.
For the $L$-wedge a similar procedure can be followed. As a result, the use of a diagonal 
vierbein in Rindler space must be accompanied by a corresponding local Lorentz 
transformation on spinor fields; in the present case this takes the form
\be
\lb' = e^{a\tau \gam_3/2} \lb.
\label{a.14}
\ee
This explains why in the $R$- and $L$-wedges of Rindler space the redefinition 
(\ref{3.9}) of the spinor fields by
\[
\ba{l}
R: \hs{2} \psi_{\pm}(\tau,\xi) = e^{\pm a (\tau \pm \xi)/2} \lb_{\pm}[t(\tau,\xi), x(\tau,\xi)], \\
 \\
L: \hs{2} \psi_{\pm}(\tau,\xi) = e^{\mp a (\tau \pm \xi)/2} \lb_{\pm}[t(\tau,\xi), x(\tau,\xi)],
\ea
\]
reduces the Rindler Dirac-Majorana action formally to the Minkowski one. 
\vs{2}

\nit
{\bf \large Appendix B: Thermal correlations in Rindler space}
\vs{1}

\nit
The computation of the two-point correlation functions of the Rindler fields in 
the Minkowski vacuum start from equations (\ref{4.5}) and (\ref{4.9}):
\be
\ba{ll}
b_{R\kg} | 0 \rangle_b = e^{-\pi\kg} b^*_{L\kg} | 0 \rangle_b, & 
b^*_{R\kg} | 0 \rangle_b = e^{\pi\kg} b_{L\kg} | 0 \rangle_b, \\
 & \\
\bg_{R\kg} | 0 \rangle_f = - i e^{-\pi\kg} \bg^*_{L\kg} | 0 \rangle_f, &
\bg^*_{R\kg} | 0 \rangle_f = i e^{\pi\kg} \bg_{L\kg} | 0 \rangle_f, 
\ea
\label{b.1}
\ee
and their conjugates:
\be
\ba{ll}
{_b\langle} 0| b^*_{R\kg} = e^{-\pi\kg} {_b\langle} 0| b_{L\kg}, & 
{_b\langle} 0 | b_{R\kg} = e^{\pi\kg} {_b\langle} 0| b^*_{L\kg}, \\
 & \\
{_f\langle} 0 | \bg^*_{R\kg} =  i e^{-\pi\kg} {_f\langle} 0| \bg_{L\kg}, &
{_f\langle} 0 |\bg_{R\kg} = - i e^{\pi\kg} {_f\langle} 0 | \bg^*_{L\kg}. 
\ea
\label{b.2}
\ee
The following chain of commutation relations and substitutions leads to the desired 
result: 
\be
\ba{lll}
{_b\langle} 0 | b_{R\kg} b^*_{R\sg} | 0 \rangle_b & = & 
 2\pi \del(\kg-\sg) + {_b\langle} 0 | b^*_{R\sg} b_{R\kg} | 0 \rangle_b \\
 & & \\
 & = & 2\pi \del(\kg-\sg) + e^{-\pi (\kg+\sg)} {_b\langle} 0 | b_{L\sg} b^*_{L\kg} | 0 \rangle_b \\
 & & \\
 & = & 2\pi \del(\kg-\sg) \lh 1 + e^{-2\pi \kg} \rh + e^{-\pi (\kg+\sg)} {_b\langle} 0 | b^*_{L\kg} b_{L\sg} | 0 \rangle_b \\
 & & \\
 & = & 2\pi \del(\kg-\sg) \lh 1 + e^{-2\pi \kg} \rh + 
  e^{-2\pi (\kg+\sg)} {_b\langle} 0 | b_{R\kg} b^*_{R\sg} | 0 \rangle_b.
\ea
\label{b.3}
\ee
The solution of this equation is (\ref{4.11}) and implies (\ref{4.12}):
\be
\ba{l}
\dsp{ {_b\langle} 0 | b_{R\kg} b^*_{R\sg} | 0 \rangle_b = 2 \pi \del(\kg - \sg) + 
 {_b\langle} 0 | b^*_{R\sg} b_{R\kg} | 0 \rangle_b = \frac{2\pi \del(\kg-\sg)}{1 - e^{-2\pi \kg}}, }\\
  \\
\dsp{  {_b\langle} 0 | b^*_{R\sg} b_{R\kg} | 0 \rangle_b = \frac{2\pi \del(\kg-\sg)}{e^{2\pi \kg} - 1}. }
\ea
\label{b.4}
\ee
The derivation of the fermion two-point function proceeds entirely parallel, using anti-commutation
relations instead of commutation relations, which results in a change of sign in the denominator.

The relations (\ref{b.1}), (\ref{b.2}) also imply a direct relation between the common Rindler vacuum 
state $| 0 \rangle_R = | 0 \rangle_{RR} \otimes | 0 \rangle_{RL}$:
\be
b_{R\kg} | 0 \rangle_{RR} = \bg_{R\kg} | 0 \rangle_{RR} = 0, \hs{2} 
b_{L\kg} | 0 \rangle_{RL} = \bg_{L\kg} | 0 \rangle_{RL} = 0,
\label{b.5}
\ee
and the Minkowski vacuum: 
\be
| 0 \rangle_M = \exp \left[ \int_0^{\infty} \frac{d\kg}{2\pi}\, e^{- \pi \kg} \lh b^*_{R\kg} b^*_{L\kg} 
 - i \bg^*_{R\kg} \bg^*_{L\kg} \rh \right] | 0 \rangle_R.
\label{b.6}
\ee

\np

\end{document}